\documentclass[final,3p,times,onecolumn]{elsarticle}
\usepackage{amsmath,amssymb,amsfonts,scalerel}
\usepackage{blindtext}
\usepackage{graphicx}
\usepackage[T1]{fontenc}
\usepackage{adjustbox}
\usepackage{epsfig}
\usepackage{float}
\usepackage{subfig}
\usepackage{xcolor}
\usepackage[colorlinks=true,citecolor=blue,linkcolor=blue,urlcolor=blue]{hyperref}
\usepackage{setspace}
\usepackage{multirow}
\usepackage{multicol}
\usepackage{url}
\usepackage{siunitx}
\usepackage{hyperref}
\usepackage{rotating}
\usepackage{geometry}
\biboptions{sort,compress}
\journal{Materialia}

\begin{document}

\begin{frontmatter}

\title{Electronic and Optical properties of TM-doped 
       (8,0)\ SiC\ SWNT and the prospect of hydrogen
       storage}
       
\author[label1,label2]{A.T.~Mulatu}
\ead{abebetadesse2@gmail.com} 

\author[label1]{K.N.~Nigussa\corref{cor1}}
\cortext[cor1]{Corresponding author:\ kenate.nemera@aau.edu.et\ 
               (K.N.~Nigussa)}
\author[label1]{L.D.~Deja}
\ead{lemi.demeyu@aau.edu.et}

\address[label1]{Department\ of\ Physics,\ Addis\ Ababa\ University,\ 
                 P.\ O.\ box\ 1176,\ Addis\ Ababa,\ Ethiopia}
\address[label2]{Department\ of\ Physics,\ Mekelle\ University,\
                 Mekelle,\ Ethiopia}
                  
\begin{abstract}
Properties of transition metal (TM)\ doped single 
wall (8,0)\ SiC\ nanotube\ is\ investigated\ 
using\ first\ principles\ density\ functional\ 
theory\ as\ implemented\ within\ quantum\ 
espresso\ code.\ The\ properties\ studied\ 
are\ electronic,\ optical,\ and\ hydrogen\ 
storage\ prospect,\ while\ the\ transition\ 
metals\ used\ in\ the\ doping\ are\ 
Iron,\ Manganese,\ and\ Cobalt.\ The\ outcomes\ 
show\ that\ ferromagnetic\ ordering\ 
better\ describes\ the\ magnetic\ order\ 
within\ the\ doping\ process.\ The\ dopings\ 
result\ in\ half-metallic\ property.\ 
Hybridization between\ TM-$3d$ and C-$2p$\
near Fermi-level\ region contributes\ 
to occurrence\ of the half-metallicity\
property.\ 
In\ addition,\ optoelectronics\ character\
and\ hydrogen\ storage\ capacity\ of\ the\ 
tube\ have\ appeared\ to\ be\ changed\ 
compared\ to\ that\ of\ the\ pristine.\   
Dopings\ have\ appeared\ to\ result\ 
in\ an\ expanded\ range\ of optoelectronics 
applications ranging from photovoltaic\ 
effects\ of far\ infrared to\ visible\ 
lights.\ Furthermore,\ the\ tube\ appears\
to\ show\ a\ potential\ for\ hydrogen\ 
energy\ storage.\  
 
\end{abstract}


\end{frontmatter}

\section{Introduction\label{sec:intro}}

A\ scientific\ pursuit\ for\ investigating\ 
efficient\ materials\ for\ optoelectronic 
analysis~\cite{WXZL2017,VB2021} 
and\ hydrogen\ storage\ applications\ 
is\ continuing\ unabated~\cite{WL2011,ASMSTPC2021}.\ 
Extracting\ energies\ using\ photovoltaic\ 
effect\ and\ from\ hydrogen\ energy\ 
is\ recently\ getting\ increased\ attention\ 
since\ it\ adheres\ to\ global\ priority\ 
area\ of\ searching\ for\ renewable\ 
energies.\ A\ range\ of\ applications\ 
such as\ spintronics\ are\ exhibited\ 
from\ half-metallicity\ properties~\cite{BTDDYDA2014,DMSBTD2015}.\ 
Literature\ review~\cite{WL2011}\ show\ that\ 
studies\ on\ Alkali-metal\ doped\ SiC\ 
nanotube\ revealed\ some\ prospect\ of\ 
hydrogen\ energy\ storage.\ Doping\ of\ 
Fe\ is\ studied\ with\ various\ systems\ 
such\ as\ ZnO~\cite{SQZC2016},\ and\ doping\ 
of\ Mn\ has\ been\ studied\ with\ 
GaP~\cite{DMSBTD2015}.\ However,\ up\ to\ date,\ 
the\ authors\ do\ not\ find\ sufficient\ 
literature\ document\ on\ doping\ of\ 
SiC\ nanotube\ with\ Fe,\ Mn,\ and\ 
Co,\ and\ the\ respective\ impacts\ 
on\ the\ electronic,\ optical,\ 
and\ hydrogen\ storage\ properties.\ 
Zigzag\ SiC\ SWNT\ show\ a direct\ 
bandgap\ in\ a\ semiconductor\ 
region~\cite{MND2021}\ while\ other\ 
chiralities\ such\ as\ armchair\ SWNT\ 
may,\ however,\ show\ indirect\ 
bandgaps~\cite{VB2021}.\ 
Thus,\ we\ believe\ that\ doping\ zigzag\ 
SiC\ SWNT\ may\ be\ better\ suitable\ 
for\ bandgap\ tuning\ for\ applications\ 
in\ half-metallicity\ and\ optoelectronics.\\

Silicon\ bulk\ system\ has\ been\ known\ 
for\ its\ good\ applications\ in\ 
photovoltaics~\cite{ASZZA2019}\ due\ its\ 
bandgap\ of\ around\ 1.1~eV~\cite{MSGLSB94}.\ 
Meanwhile,\ carbon-graphite\ is\ known\ for\ 
its\ electrical conductor\ nature\ with\ 
narrow\ or no\ bandgap~\cite{ZWWW2009}.\ 
Silicon\ carbide\ bulk\ is\ expected\ to\ 
have\ bandgap\ of\ 2.2~eV~\cite{MSGLSB94}.\ 
In\ our\ previous\ work~\cite{MND2021},\ 
we\ have\ reported\ that\ nanotubes\ have\ 
bandgaps\ smaller\ than\ the\ bulk\ 
counterpart\ but\ where\ the\ bandgaps\ 
approach\ closely\ to\ that\ of\ the\ bulk\ 
value\ for\ zigzag\ chiralities\ greater\ 
or\ equal\ to\ (6,0).\ By\ doping,\ 
the\ bandgaps\ would\ be\ likely\ 
to\ be\ reduced\ from\ the\ value\ 
corresponding\ to\ the\ pristine,\ 
and\ thus\ would\ imply\ different\ 
electrical\ as\ well\ as\ optoelectronics\ 
application\ properties.\\

Dilute\ magnetic\ semiconductor~(DMS)\ 
property\ is\ reported\ to\ be\ exhibited\ 
by\ doping\ semiconductors\ with\ transition\ 
metals~\cite{ZHLLT2020,SYD2003}.\ 
DMSs often\ exhibit\ half-metallic\ property\
and\ can\ undergo\ a transition\ from\ 
ferromagnetic\ state\ to\ paramagnetic\ 
state\ at a temperature\ called\ critical\ 
temperature,\ which is\ calculated\ by\ 
\begin{equation}{\label{eq1}}
T\rm_{c}=\frac{2{\Delta}E\rm_{FM}}{3{k\rm_{B}}n}
\end{equation} 
as\ described\ in\ literature~\cite{ZHLLT2020},
where\ $k\rm_{B}$ is the Boltzmann constant,\ 
$\rm{n}$ is the number of doping atoms,\
and ${\Delta}E\rm_{FM}$~is\ the\ difference\ 
between\ total energies in the\ ferromagnetic 
and antiferromagnetic ordering.\ 
The\ prospect\ of\ occurrence\ 
of\ such\ property\ in\ (8,0) SiC SWNT\ 
is\ studied\ by\ doping\ with\ Fe,\ 
Mn,\ $\&$\ Co.\ Dopant concentration\ 
$x$ is\ described as\ the fraction ratio\
of number\ of dopant atoms\ to number\ 
of Silicon atoms\ in the unit cell\ 
of\ the pristine.\ Thus,\ $x=0.0625$, \
and $x=0.125$ are considered in this 
study.\ With\ consideration\ of\ the\ 
dopant\ atoms\ Fe,\ Mn,\ and Co,\ 
we expect\ to\ investigate\ the\ 
relative\ prospects\ of\ applications\ 
in\ spin\ injection\ as well\ as\ at\ 
which\ concentartions\ do\ such\ applications\ 
become\ optimal.\ The paper is organized\ 
as\ follows.~In the\ next\ section\
(sec.~\ref{sec:comp}),\ a\ detail account\ 
of the computational\ method\ is presented.\ 
Results\ and discussion are\ presented in\ 
section~\ref{sec:res},\ with\ the conclusion\ 
being presented in\ section~\ref{sec:conc}.
\section{Computational Methods\label{sec:comp}}

An ab-initio simulations within quantum 
espresso\ code~\cite{Giannozzietal2009}\
is\ used\ to examine the\ electronic structure\ 
and optical properties of the\ zigzag\ 
(8,0)~SiC\ SWNT.\ The electron wavefunction\ 
is expanded over\ a plane wave basis\ 
set.\ The\ electron-ion\ interactions\ is\ 
approximated\ within\ projector augmented\ 
wave~(PAW) modality\ (PAW data set)~\cite{PB94}\ 
upon\ the\ calculation\ of\ electronic\ properties\ 
and geometry\ optimizations.\ Upon\ optical\ 
properties\ calculations,\ the\ electron-ion\ 
interactions\ is\ approximated\ within\ norm\ 
conserving\ pseudopotential~\cite{KB82}.\  
The exchange-correlation\ energies are\ treated\ 
using\ PBE~\cite{PBE96}.\
The\ k-points\ of\ the\ Brillouin zone~(BZ)\ are\ 
generated from the input ${\bf{k}}$-mesh\ using\ 
the\ Monkhorst-Pack scheme~\cite{MP76}.\
The\ number\ of valence electrons\ considered\ for\ 
each element\ within the\ paw\ data\ sets\ is\ Fe:8,\ 
Mn:7,\ Co:9,\ H:1, C:4,\ and Si:4.\ Geometry\ 
relaxations are\ carried out using BFGS\ minimizer~\cite{BS82},\ 
where optimizations\ of\ the atomic\ coordinates\ 
and the unit\ cell degrees of freedom\ is done\ 
within\ the concept\ of the Hellmann-Feynman forces\ 
and\ stresses~\cite{PRF39, NM85}\ as\ calculated on\ 
the Born-Oppenheimer~(BO) surface~\cite{WM91}.\ 
The convergence\ criteria for the forces\ were set\ 
at 0.05 eV/{\AA}.

The\ optimum\ magnetic\ moments\ of\ 4.0~${\mu}\rm_{B}$,\
3.0~${\mu}\rm_{B}$,\ and 3.0~${\mu}\rm_{B}$,\
is\ placed\ on\ each Fe,\ Mn,\ $\&$\ Co\ atoms,\ 
respectively.\ 
Dopings\ is\ done\ by\ replacing\ two\ or\ 
four\ silicon\ atoms\ with\ the\ dopant\ 
transition\ metal\ atoms.\ The\ favored\ 
magnetic\ ordering\ is\ ferromagnetic.\ 
When the number of dopant atoms is two, 
the\ total\ magnetic\ moment\ per\ unit\ 
cell\ 8.0~${\mu}\rm_{B}$,\ 6.0~${\mu}\rm_{B}$,\
6.0~${\mu}\rm_{B}$,\ respectively, for Fe,
Mn,\ and Co.\ When the number of dopant\ 
atoms is four,\ the\ total\ magnetic\ moment\ 
per\ unit\ cell\ 16.0~${\mu}\rm_{B}$,\ 
12.0~${\mu}\rm_{B}$,\ 12.0~${\mu}\rm_{B}$,\ 
respectively,\ for Fe,\ Mn,\ and Co.\ 
The\ pristine\ nanotube\ unit\ cell\ contains\ 
64\ atoms\ with Si:C ratio being 1:1,\ and\ has\ a 
tetragonal\ lattice\ vectors\ $a=b=15.9~{\AA}$\ 
and\ $c=10.8~{\AA}$,\ with\ periodic\ boundary\ 
conditions\ being\ allowed\ along\ the\ $c$-lattice\ 
vector,\ and\ a\ zero\ chirality\ angle.\ Along\ 
the\ $a~\&~b$\ lattice\ vectors,\ 
a\ vacuum\ space\ of\ 10~{\AA}\ is\ placed\ to\ 
make\ sure\ that\ no\ lateral\ interaction\ 
occurs\ between\ adjacent\ unit\ cells.\\

Formation\ energy~(eV/atom)~is\ calculated\ 
as\ 
\begin{equation}{\label{eq2}}
E_{f}={{\mu}\rm_{t}}-{{\mu}\rm_{Si}}-{{\mu}\rm_{C}}
-{{\mu}\rm_{dopant}}
\end{equation}      
where\ ${{\mu}\rm_{t}}$\ is\ total\ chemical\ 
potential\ of\ the\ system\ containing\ 
dopant\ and\ tube,\ ${{\mu}\rm_{Si}}$\ is\ 
chemical\ potential\ of\ bulk\ Silicon\ 
crystal,\ ${{\mu}\rm_{C}}$\ is\ 
chemical\ potential\ of\ bulk\ Carbon\ 
crystal\ in\ graphite\ structure,\ 
$\&$\ ${{\mu}\rm_{dopant}}$\ is\ chemical\ 
potential\ of\ the\ dopant\ atom.\ Iron,\ 
Manganese,\ and Cobalt\ are\ dopant\ atoms\ 
considered\ in\ this\ work.\ Chemical\ 
potentials\ are\ calculated\ as\ total\ 
energy\ divided\ by\ the\ number\ of\ 
atoms\ in\ the\ unit\ cell\ of\ the\ 
bulk\ crystals,\ while\ for\ the\ nanotube,\ 
it\ is\ given\ by\ total\ energy\ divided\ 
by\ number\ of\ SiC\ pairs,\ $\&$\ 
for\ dopant\ atoms\ it\ is\ the\ same\ as\ 
total\ energy\ of\ single\ free\ atom.\ Free\ 
atom\ is\ modelled\ by\ putting\ single\ 
atom\ in\ a\ cubic\ super\ unit\ cell\ of\ 
side\ length~10~{\AA}.\ 
   
Adsorptions\ are\ done\ by\ putting\ adsorbate\ 
molecules\ on\ the\ tube\ $\&$\ then\ allowing\ 
geometry\ relaxations\ to\ take\ place.\
Adsorption\ energy\ is\ calculated\ as\
\begin{equation}{\label{eq3}}
E_{ads}=-({E\rm_{(t~+~nH_{2})}}-{E\rm_{t}}-{E\rm_{nH_{2}}})
\end{equation}
where\ ${E\rm_{(t~+~nH_{2})}}$\ is\ total\ 
energy\ of\ the\ system\ containing\ 
tube~$\&$~adsorbate\ hydrogen,\
${E\rm_{t}}$\ is\ total\ energy\ of\ 
clean\ tube,~$\&$~$n{E\rm_{H_{2}}}$\ is\ 
total\ energy\ of\ adsorbate\ hydrogen.\ 
${E\rm_{H_{2}}}$\ is\ total\ energy\ of\ 
a\ single\ free\ hydrogen\ molecule,\ which\ 
is\ obtained\ by\ optimizing\ in\ a\ cubic\ 
supercell\ of\ side\ length\ 10~{\AA}.\ 
The\ number\ of\ adsorbate\ molecules\ 
used\ in\ the\ adsorption\ is\ indicated\ 
by\ $n$.\ According\ to\ Eq.~\eqref{eq3},\ 
positive\ values\ of\ $E\rm_{ads}$\ means\ 
exothermic\ process,\ while\ negative\ values\ 
mean\ endothermic.\ A\ van\ der\ Waal's\ 
treatment\ within\ DFT-D3~\cite{GEG2011}\ 
is\ applied\ wherever\ necessary.\ 
The {\bf{k}}-mesh of 1{$\times$}1{$\times$}15\ 
and a\ cut-off\ energy~(ecut)~of\ 600 eV is\ 
used\ in\ the\ calculations.\

Hubbard U correction~\cite{AZA91}\ is\ applied\ 
to\ the dopant atoms.~We\ have selected\ U=7~eV\ 
to be optimum to our system.~Spin polarized 
calculation is allowed.~Density\ of\ states~(DOS)\ 
is\ calculated as a\ population\ of states\ 
in the spin-up and spin-down states\ at\ 
the\ chosen\ energy\ windows.~Projected\ 
DOS~(PDOS)\ is\ calculated as a\ component\ 
of\ DOS\ resolved\ onto atomic\ orbitals.\
To\ characterize\ optical\ properties,\ 
a\ dielectric\ function\ is\ computed,\ which\ 
has\ generally\ a\ complex\ nature\ $\&$\ 
is\ given\ as
\begin{equation}{\label{eq4}}
\varepsilon(\omega)={\varepsilon\rm_{1}}(\omega)\ +\ i\ 
{\varepsilon\rm_{2}}(\omega)
\end{equation}
The imaginary part\ ${\varepsilon\rm_{2}}(\omega)$\ 
is\ calculated\ from the density matrix of the\ 
electronic\ structure~\cite{HL87}\ as\ described\ 
elsewhere~\cite{GHKFB2006},\ $\&$\ given\ 
by\
\begin{equation}{\label{eq5}}
{\varepsilon\rm_{2}}(\omega)=\frac{8{{\pi}^{2}}e^{2}{\hbar}^{2}}
{\Omega {{\omega}^{2}}{m_{e}}^{2}} 
{\sum\limits_{k,v,c}}{w\rm_{k}}{{\mid}\langle{\psi\rm_{k}^{c}}
{\mid}{\bf u}{\cdot}
{\bf r}{\mid}{\psi\rm_{k}^{v}}\rangle{\mid}}^{2}
\delta(E\rm_{k}^{c}-E\rm_{k}^{v}-\hbar \omega), 
\end{equation}
where $e$ is the electronic charge,\ and $\psi\rm_{k}^{c}$\ 
and\ $\psi\rm_{k}^{v}$\ are the conduction band\ 
(CB)\ and\ valence band\ (VB)\ wave functions at k,\ 
respectively,\ $\hbar \omega$\ is the\ energy of the\ 
incident phonon,\ ${\bf u}{\cdot}{\bf r}$ is\ the\ 
momentum operator,\ $w\rm_{k}$\ is a joint\ density\ 
of states,\ $\&$\ $\Omega$\ is\ volume\ of\ the\ 
primitive cell.\ The real\ part\ ${\varepsilon\rm_{1}}(\omega)$\ 
can\ be extracted from the\ imaginary part of\ 
Eq.~\eqref{eq5}\ according to Kramer-Kronig\ 
relationship~\cite{FW72},\ as follows.\
\begin{equation}{\label{eq6}}
{\varepsilon\rm_{1}}(\omega)=1\ +\ {\frac{2}{\pi}}P
{\int\limits_{0}^{\infty}}
\frac{{\omega}'{\varepsilon\rm_{2}}({\omega}')}
{{{\omega}'}^{2}-{{\omega}}^{2}}d{\omega}'
\end{equation}
where\ $P$\ is\ a\ principal\ value.\ 
The electron energy loss function~($L({\omega})$),\
as\ given\ elsewhere~\cite{SSM2000},~is\ 
calculated~by\
\begin{equation}{\label{eq7}}
\frac{{\varepsilon\rm_{2}(\omega)}}
{{\varepsilon\rm_{1}^{2}(\omega)} 
+ {\varepsilon\rm_{2}^{2}(\omega)}}
\end{equation}
The joint density of\ states~(JDOS)\ 
is defined as\
\begin{equation}{\label{eq8}}
n(\omega)=\sum_{\sigma}\sum_{n{\in}V}\sum_{n'{\in}C}
\frac{\Omega}{(2\pi)^{3}}{\int} 
{\delta}(E_{k,n'}-E_{k,n}-{\hbar}{\omega}){d^{3}\textbf{k}}
\end{equation}
were $\sigma$ is the spin component,\ $\Omega$\ 
is the volume of the\ lattice cell,\ $n$ and $n'$\ 
belong\ to the valence\ and conduction\ 
bands,\ respectively,\ and $E_{k,n}$ and\ 
$E_{k,n'}$\ are the\ corresponding\ 
eigenvalues.\
\section{Results and discussion\label{sec:res}}
\begin{table}
\setlength{\tabcolsep}{6.0mm}
\renewcommand{\arraystretch}{1.5}
\centering
\small
\caption{Concentration\ of dopant~($x$),\ Half-metallic\ 
         energy\ gap~($HM\rm_{g}$)~in~eV,\ Formation\ 
         energy~($E\rm_{f}$)~in~eV/atom,\ Total energy per\ 
         atom~($E$)~in~eV/atom,\ Energy\ 
         gap~($E\rm_{g}$)~in~eV,\ and critical\ 
         temperature $T\rm_{c}$ in Kelvin,\ for\ 
         doped and pristine\ (8,0)\ SiC~SWNT.\label{tab1}}
\begin{tabular}{lcccccccc}
\hline
{System} & $x$ & $E\rm_{g}$~(eV) & $E\rm_{f}$~(eV/atom) & $E$~(eV) 
& $HM\rm_{g}$~(eV) & $T\rm_{c}$~(K)\\
\hline
Pristine &  -     & 1.10 & -1.09 & -9.59  &  -   &  -     \\
\hline
         & 0.0625 & 0.80 & 3.65  & -17.05 & 0.38 &  -      \\
         & 0.1250 & 1.59 & 1.47  & -24.65 & 0.81 & 61.89   \\
Fe-doped & 0.1875 & 1.69 & 1.73  & -32.17 & 0.43 &  -      \\
         & 0.2500 & 2.12 & 0.56  & -39.74 & 0.29 & 385.29  \\
\hline
        
          & 0.1250 & 1.29 &-1.22  & -26.44 & 1.14 & 527.04   \\
Co-doped  & 0.1875 & 1.58 & 8.48  & -34.83 & 1.20 & 311.00    \\
          & 0.2500 & 1.93 & 11.20 & -43.24 & 1.36 &           \\
\hline
        
         &0.0625 & 1.14 & 5.92    & -15.87 & 0.52 &          \\
         &0.1250 & 0.88 & -113.97 & -22.45 & 0.30 & 108.31   \\
Mn-doped &0.1875 & 1.54 & -163.53 & -28.42 & 1.26 & 638.11   \\
         &0.2500 & 1.97 & 7.307   & -34.74 & 1.45 & 622.33    \\
\hline
\end{tabular}
\end{table} 
As\ can\ be\ seen\ from table~\ref{tab1},\ the formation\ 
energies\ have\ positive\ and negative values.\ According\
to\ Eq.~\eqref{eq2},\ negative values\ means exothermic,\
while positive values means endothermic\ process.\
Thus,\ the dopings somehow\ result in\ less stable\
structures\ as\ compared\ to\ the\ pristine.\ 
However,\ increasing\ concentration\ of\ dopant\ 
seems to increase\ back\ the\ stabilities\ 
of\ structures\ but\ where\ the\ latter\ 
stabilities\ are\ still\ lower\ than\ that\ 
of\ the pristine.\ 
The\ critical\ temperatures\ are\ calculated\ 
according\ to\ Eq.~\ref{eq1},\ and\ show\ increases\ 
in\ values\ when\ the\ dopant\ concentration\ 
increases.\   

\subsection{Electronic\ properties of TM doped (8,0)\ SiC~SWNT}
By\ doping\ Fe,\ the bandgap seems\ 
to increase\ with increasing\ concentration\ 
of dopant while\ half-metallic\ bandgap\ 
seems to decrease~(see table~\ref{tab1}).\ 
The\ values of $HM\rm_{g}$\ are\ well\ 
in\ a\ range\ suitable\ for\ spintronic\ 
applications\ at\ all\ dopant\ concentrations\ 
considered\ in\ this\ study.\ 
\begin{figure*}[ht!]
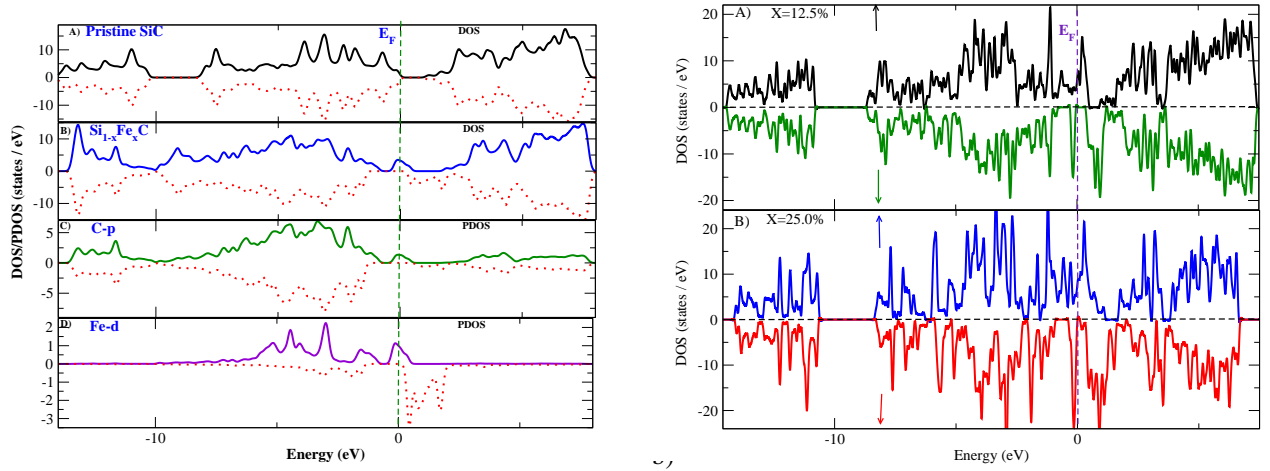

\subfloat a){{\includegraphics[width=.47\textwidth]{./Figures/Fig1a.eps}}}\qquad
\subfloat b){{\includegraphics[width=.47\textwidth]{./Figures/Fig1b.eps}}}\qquad
\caption{a)~Left side. PDOS of Si$\rm_{1-x}$Fe$\rm_{x}$C~($x = 0.0625$)\ 
           compound. b)~Right side. DOS for majority/minority\ 
           spin\ channels~(MAC/MIC)\ is shown by +ve/-ve value.\
           X\ is meant for percentage form of $x$.\ \label{fig1}}
\end{figure*}  
The\ DOS\ of\ pristine\ on\ the\ left\ 
side\ of Fig.~\ref{fig1},\ top\ part,\ 
shows\ no\ half-metallicity\ property\ 
since\ both\ spin-up and\ spin-down\ 
channels\ do\ not\ cross\ the\ Fermi-level.\ 
However,\ by\ the\ introduction\ of\ the\ 
dopant\ Fe,\ the\ DOS\ of\ MAC\ shows\ crossing\ 
of\ the\ Fermi-level\ while the MIC\ 
shows a gap~(see\ right\ side\ of\ the figure).\  
Meanwhile,\ the\ PDOS~(left side\ of Fig.~\ref{fig1},\
lower parts)\ show\ that\ the hybridization between\
Fe-$3d$\ and O-$2p$\ is\ resulting\ to\ the\ 
non-zero\ DOS\ values\ of\ the majority\ 
spin\ channel.\\ 
 
With\ Cobalt and Manganese doped SiC~SWNT,\ 
the bandgap and half-metallic\ gap seem\ 
to\ increase\ by\ increasing\ dopant\ 
concentrations~(see\ table~\ref{tab1}).\ 
Thus,\ Co\ doping\ seems\ to\ show\ 
little\ significance\ for\ application\ 
in\ spintronics,\ within\ the\ range\ 
of\ dopant\ concentration\ considered\ 
in\ this\ study.\ However,\ Manganese\ 
doping\ can\ be\ applied\ in\ spintronics\ 
application\ at\ lower\ dopant\ concentrations\ 
of\ up\ to\ $x=0.125$.\

\begin{figure*}[ht!]
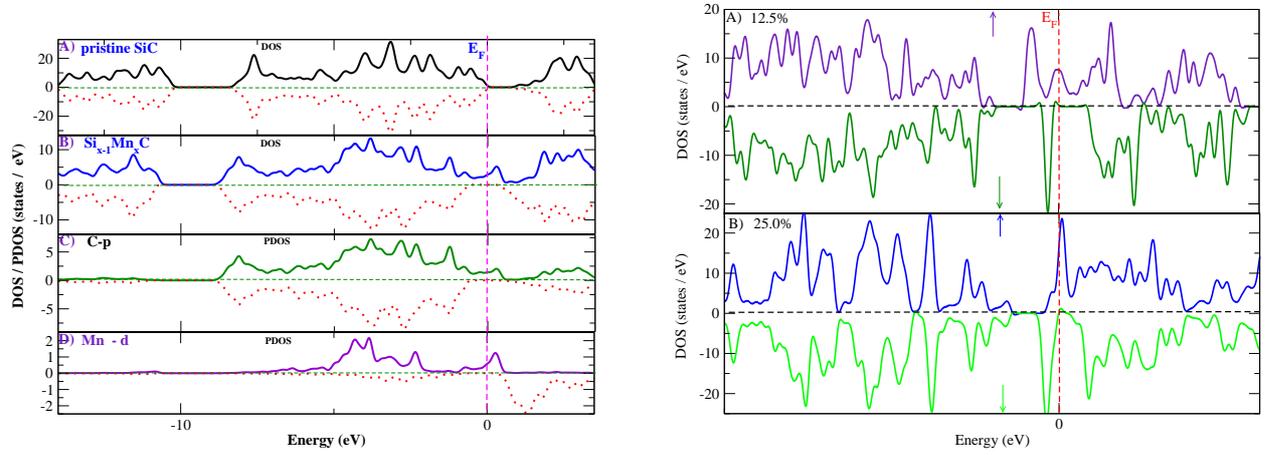

\subfloat a){{\includegraphics[width=.47\textwidth]{./Figures/Fig2a.eps}}}\qquad
\subfloat b){{\includegraphics[width=.47\textwidth]{./Figures/Fig2b.eps}}}\qquad
\caption{a)~Left side. PDOS of Si$\rm_{1-x}$Mn$\rm_{x}$C~($x = 0.0625$)\ 
           compound. b)~Right side. DOS for majority/minority\ 
           spin\ channels~(MAC/MIC)\ is shown by +ve/-ve value.\
           X\ is meant for percentage form of $x$.\ \label{fig2}}
\end{figure*}

The\ DOS\ of\ pristine\ on\ the\ left\ 
side\ of Fig.~\ref{fig2},\ top\ part,\ 
shows\ no\ half-metallicity\ property\ 
since\ both\ spin-up and\ spin-down\ 
channels\ do\ not\ cross\ the\ Fermi-level.\ 
However,\ by\ the\ introduction\ of\ the\ 
dopant\ Mn,\ the\ DOS\ of\ MAC\ shows\ crossing\ 
of\ the\ Fermi-level\ while the MIC\ 
shows a gap~(see\ right\ side\ of\ the figure).\  
Meanwhile,\ the\ PDOS~(left side\ of Fig.~\ref{fig1},\
lower parts)\ show\ that\ the hybridization between\
Mn-$3d$\ and O-$2p$\ is\ resulting\ to\ the\ 
non-zero\ DOS\ values\ of\ the majority\ 
spin\ channel.\
\begin{figure*}[ht!]
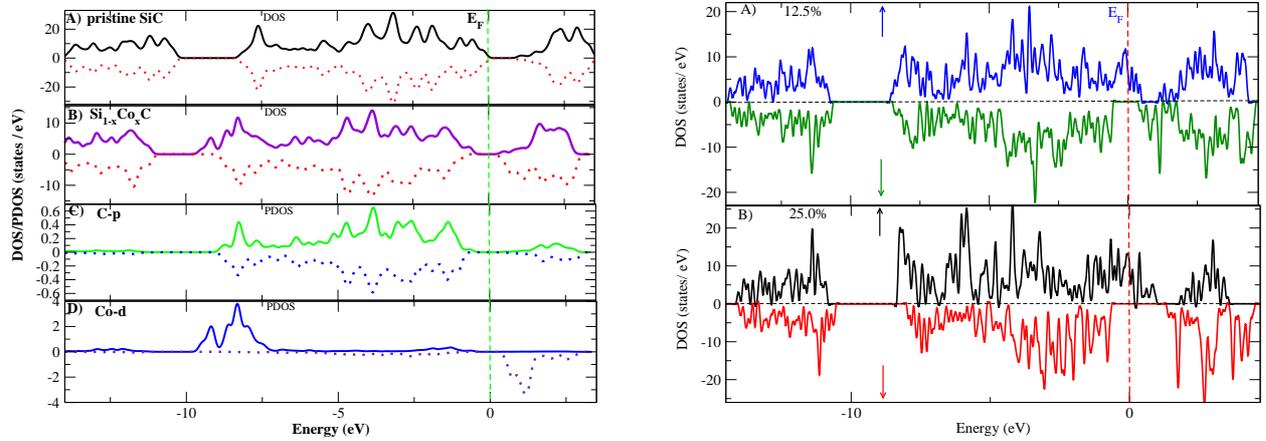

\subfloat a){{\includegraphics[width=.47\textwidth]{./Figures/Fig3a.eps}}}\qquad
\subfloat b){{\includegraphics[width=.47\textwidth]{./Figures/Fig3b.eps}}}\qquad
\caption{a)~Left side. PDOS of Si$\rm_{1-x}$Co$\rm_{x}$C~($x = 0.0625$)\ 
           compound. b)~Right side. DOS for majority/minority\ 
           spin\ channels~(MAC/MIC)\ is shown by +ve/-ve value.\
           X\ is meant for percentage form of $x$.\ \label{fig3}}
\end{figure*}
The\ DOS\ of\ pristine\ on\ the\ left\ 
side\ of Fig.~\ref{fig3},\ top\ part,\ 
shows\ no\ half-metallicity\ property\ 
since\ both\ spin-up and\ spin-down\ 
channels\ do\ not\ cross\ the\ Fermi-level.\ 
However,\ by\ the\ introduction\ of\ the\ 
dopant\ Co,\ the\ DOS\ of\ MAC\ shows\ crossing\ 
of\ the\ Fermi-level\ while the MIC\ 
shows a gap~(see\ right\ side\ of\ figure).\
However,\ this\ happens\ only\ at\ higher\ 
Co\ dopant\ concentration,\ starting at\ 
$x=0.125$.\ Meanwhile,\ the\ corresponding\ 
PDOS\ show\ that\ the hybridization between\
Co-$3d$\ and O-$2p$\ is\ resulting\ to\ the\ 
non-zero\ DOS\ values\ of\ the majority\ 
spin\ channel.\
\subsection{Optical Properties of TM doped (8,0)\ SiC~SWNT} 
\begin{figure*}[ht!]
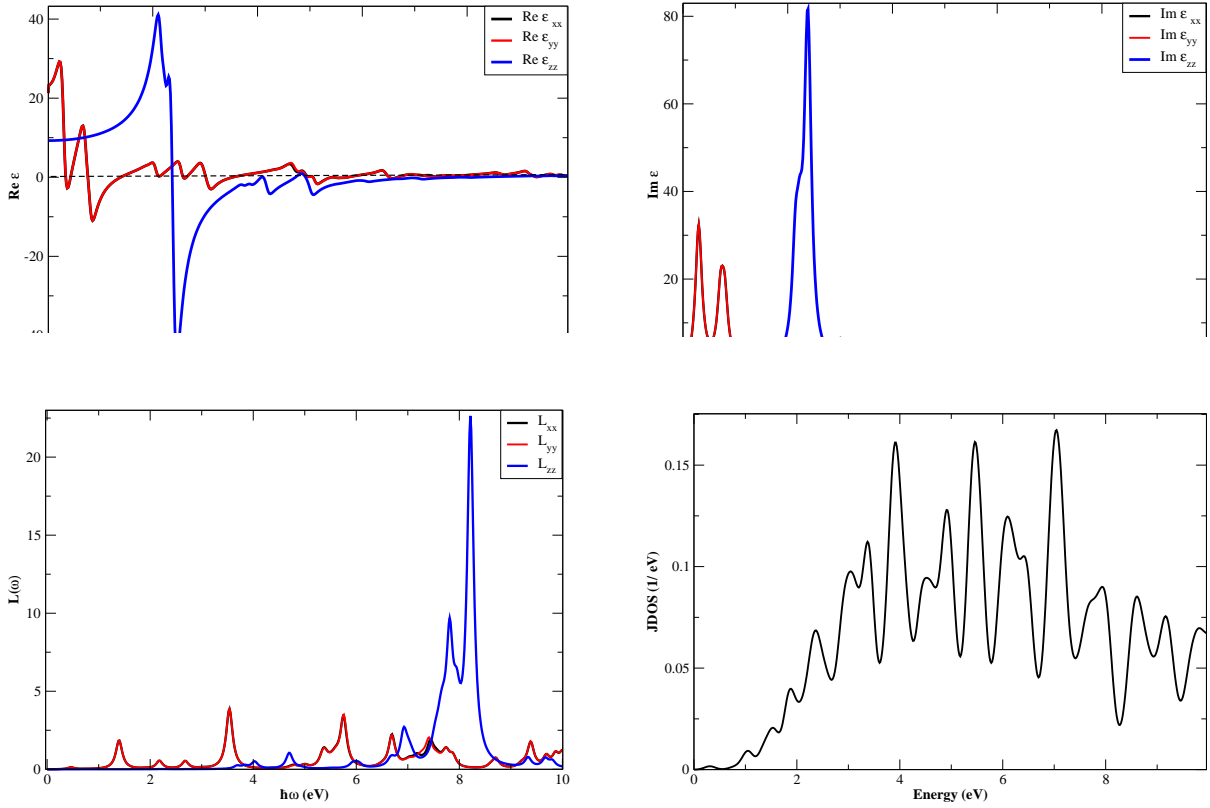

\subfloat a){{\includegraphics[width=.45\textwidth]{./Figures/Fig4a.eps}}}\qquad
\subfloat b){{\includegraphics[width=.45\textwidth]{./Figures/Fig4b.eps}}}\qquad
\subfloat c){{\includegraphics[width=.45\textwidth]{./Figures/Fig4c.eps}}}\qquad
\subfloat d){{\includegraphics[width=.45\textwidth]{./Figures/Fig4d.eps}}}\qquad
\caption{Optical\ properties\ of\ pristine\ (8,0)\
         SiC~SWNT.~a)~Top left.\ Real part\ of\ the\ 
         dielectric function\ calculated\ 
         according\ Eq.~\eqref{eq6}.~b)~Top\ right.\ 
         Imaginary\ part\ of\ dielectric function\ 
         calculated according to\ Eq.~\eqref{eq5}.\
         c)~Bottom\ left.\ Electron energy loss\ 
         calculated according to\ Eq.~\eqref{eq7}.\
         d)~Bottom right.\ JDOS\ calculated\ 
         according to\ Eq.~\eqref{eq8}.\label{fig4}}
\end{figure*}
\begin{figure*}[ht!]
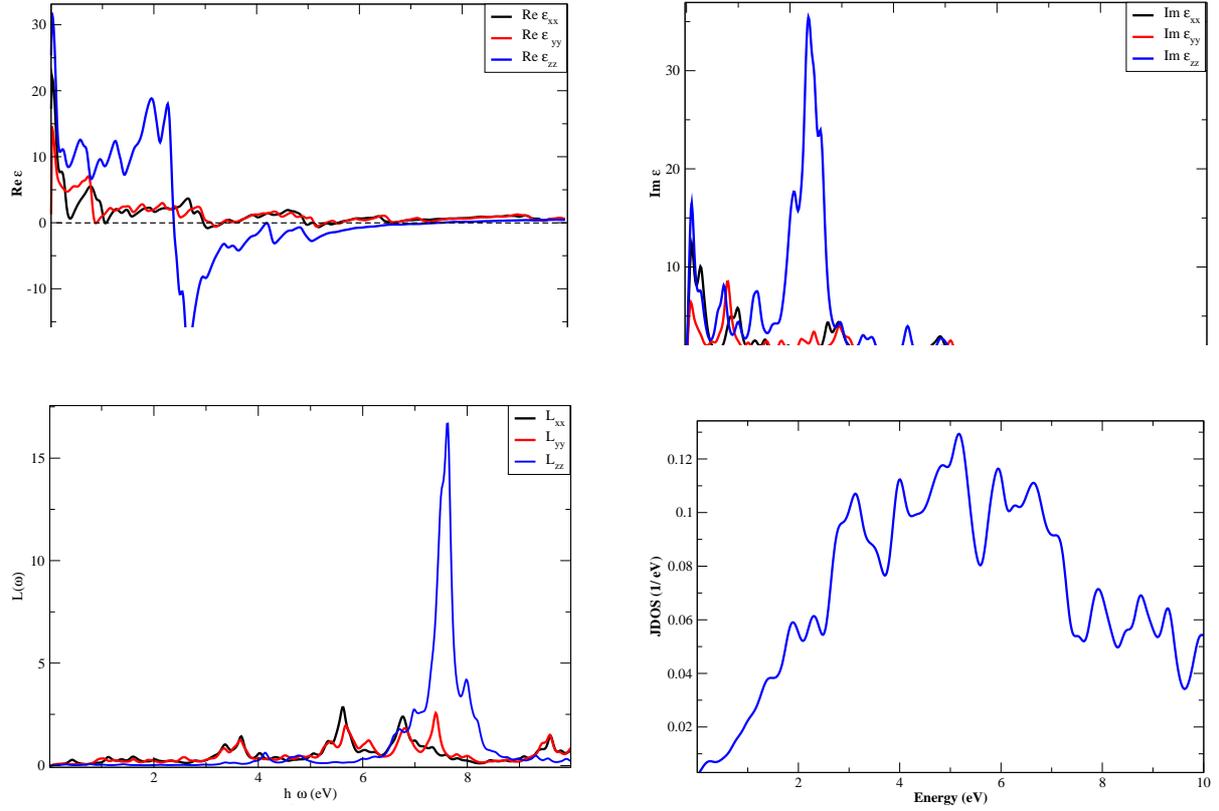

\subfloat a){{\includegraphics[width=.45\textwidth]{./Figures/Fig5a.eps}}}\qquad
\subfloat b){{\includegraphics[width=.45\textwidth]{./Figures/Fig5b.eps}}}\qquad
\subfloat c){{\includegraphics[width=.45\textwidth]{./Figures/Fig5c.eps}}}\qquad
\subfloat d){{\includegraphics[width=.45\textwidth]{./Figures/Fig5d.eps}}}\qquad
\caption{Optical\ properties\ of\ Fe-doped\ (8,0)\
         SiC~SWNT.~a)~Top left.\ Real part\ of\ the\ 
         dielectric function\ calculated\ 
         according\ Eq.~\eqref{eq6}.~b)~Top\ right.\ 
         Imaginary\ part\ of\ dielectric function\ 
         calculated according to\ Eq.~\eqref{eq5}.\
         c)~Bottom\ left.\ Electron energy loss\ 
         calculated according to\ Eq.~\eqref{eq7}.\
         d)~Bottom right.\ JDOS\ calculated\ 
         according to\ Eq.~\eqref{eq8}.\label{fig5}}

\end{figure*}
\begin{figure*}[ht!]
\subfloat a){{\includegraphics[width=.45\textwidth]{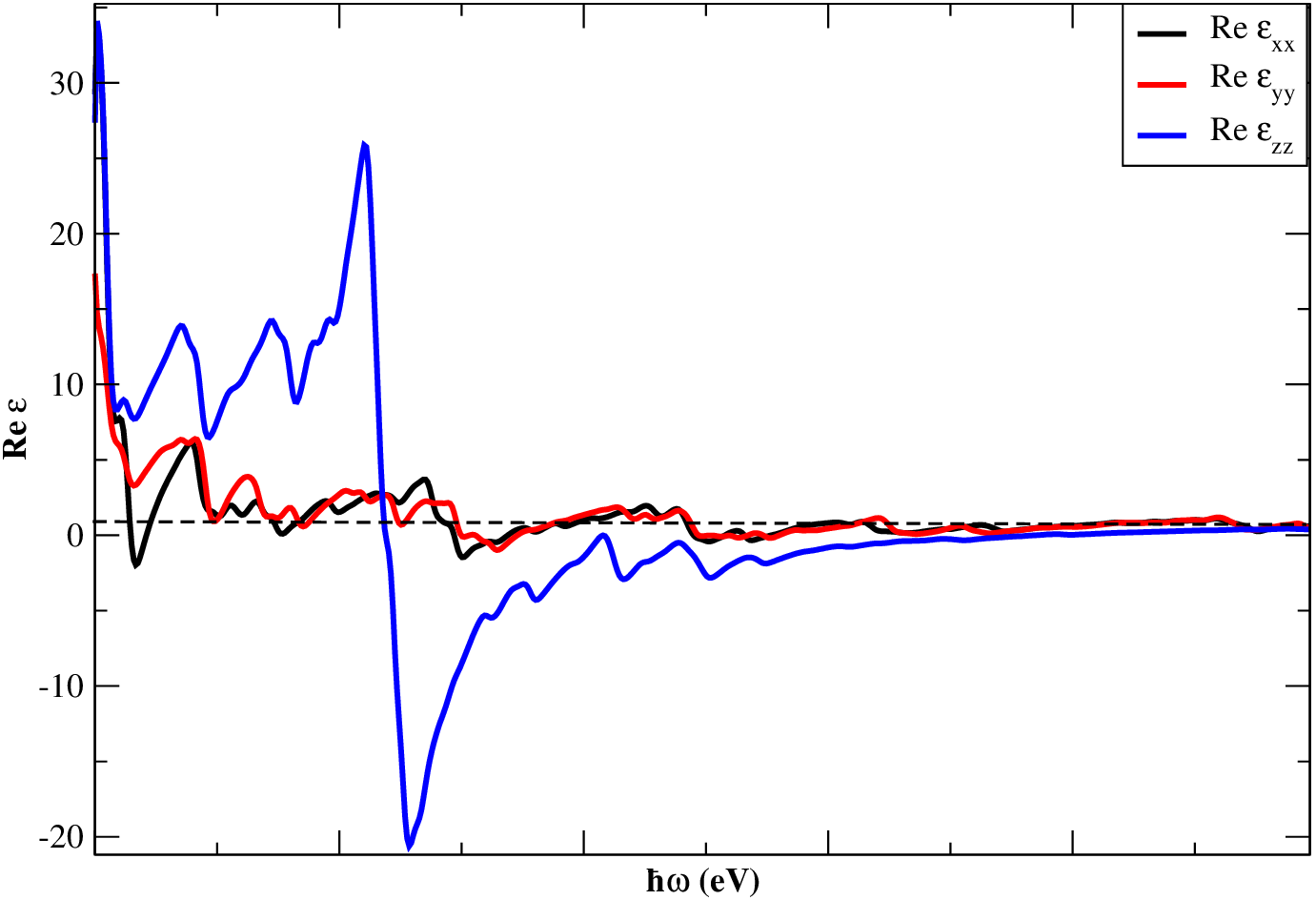}}}\qquad
\subfloat b){{\includegraphics[width=.45\textwidth]{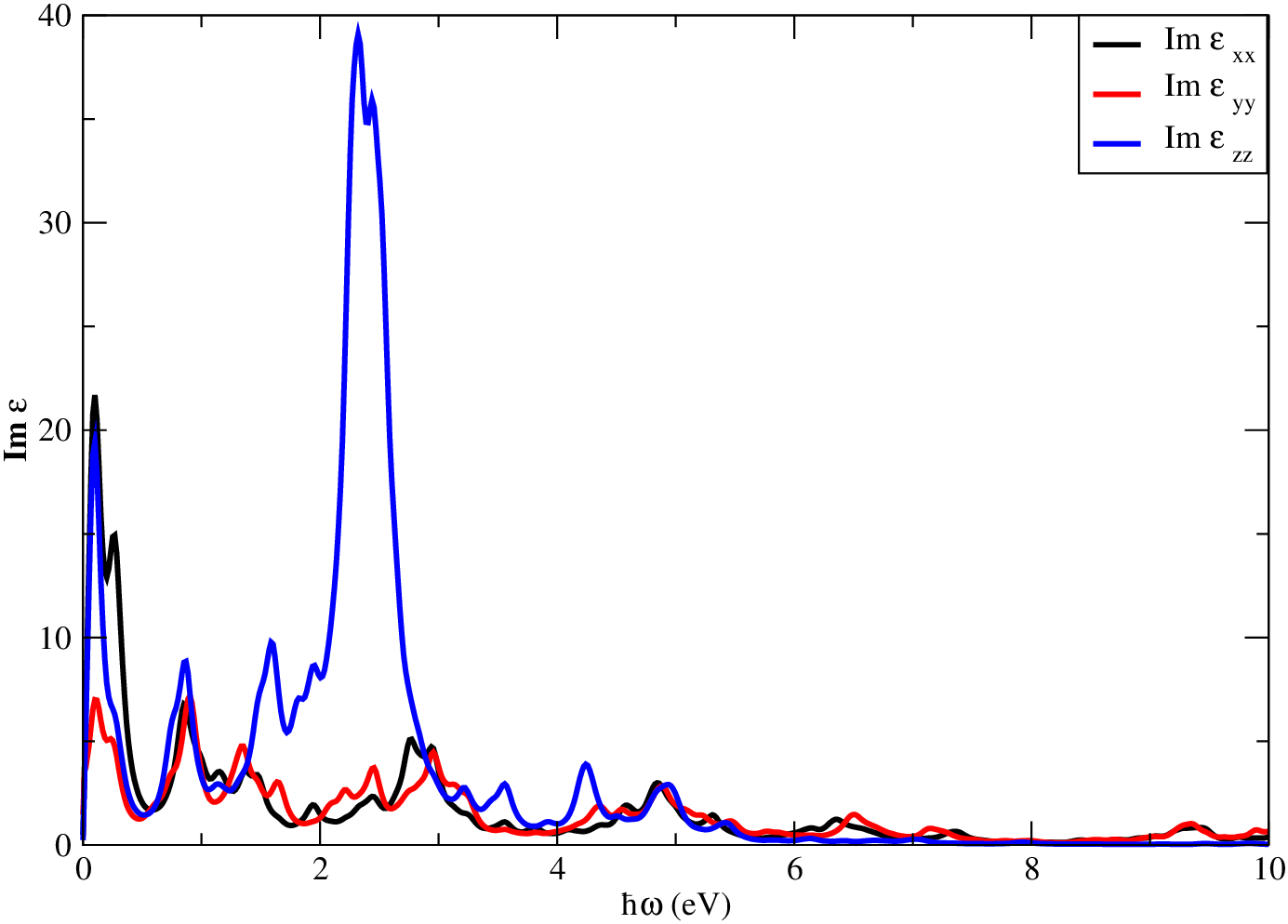}}}\qquad
\subfloat c){{\includegraphics[width=.45\textwidth]{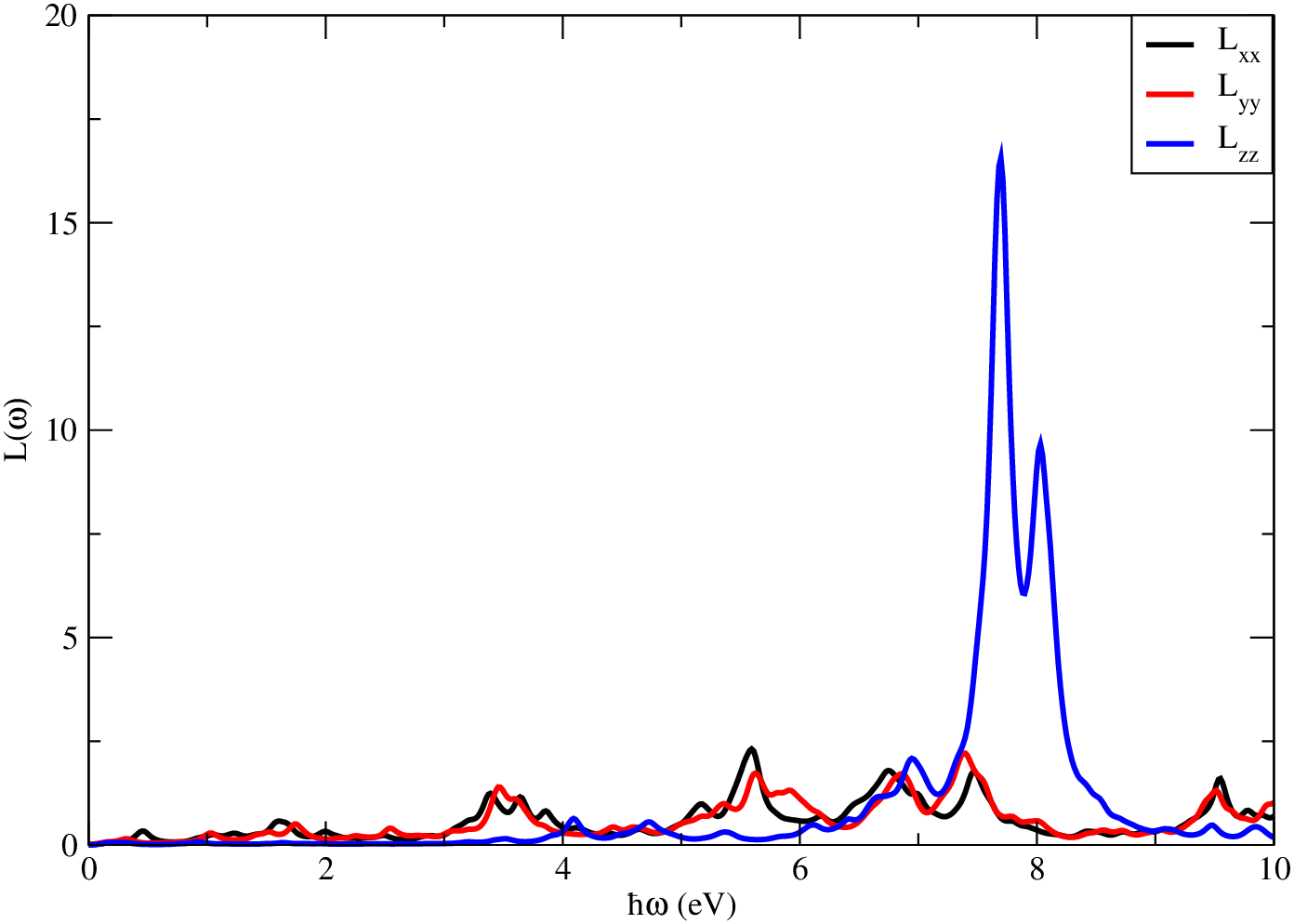}}}\qquad
\subfloat d){{\includegraphics[width=.45\textwidth]{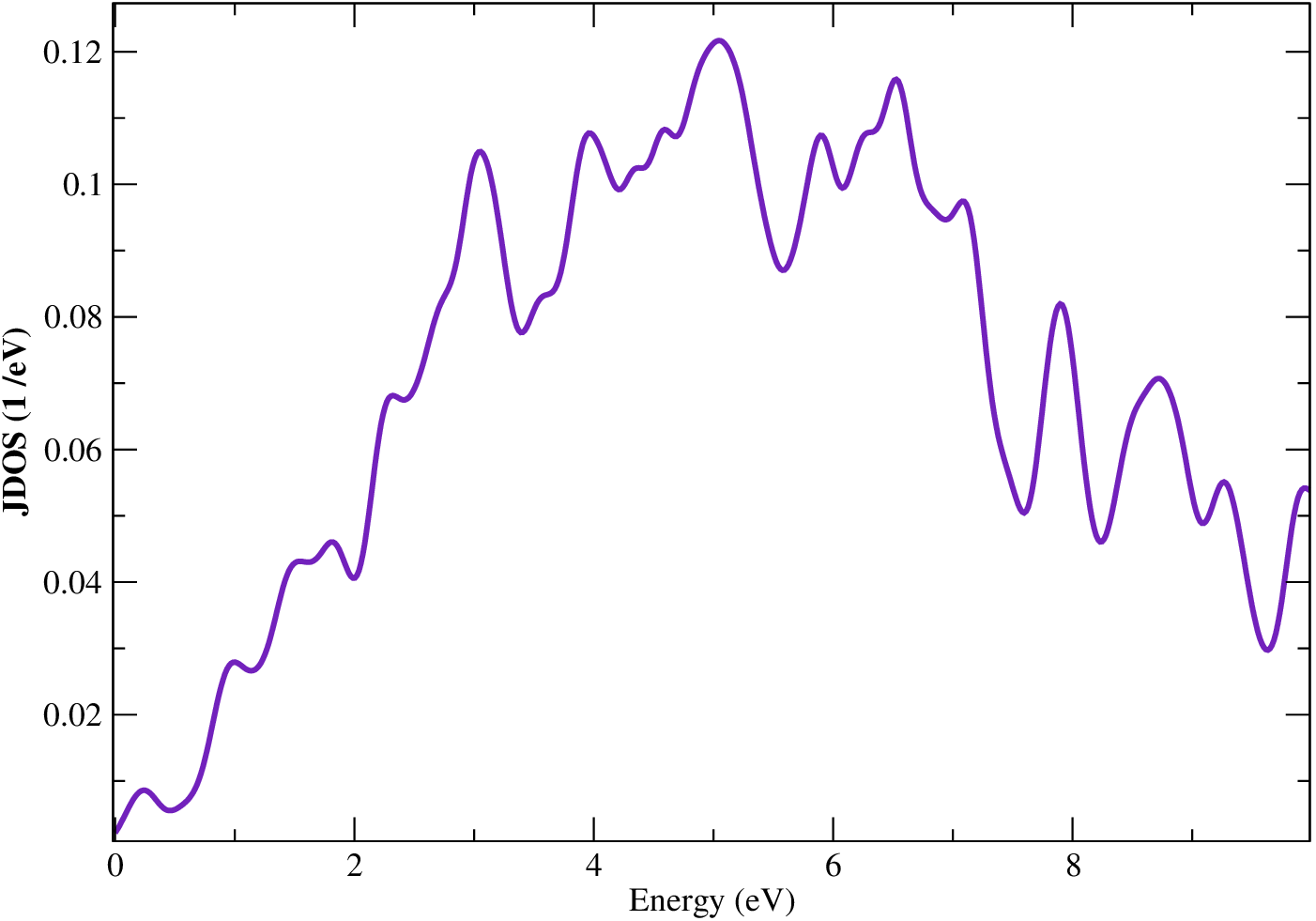}}}\qquad
\caption{Optical\ properties\ of\ Co-doped\ (8,0)\
         SiC~SWNT.~a)~Top left.\ Real part\ of\ the\ 
         dielectric function\ calculated\ 
         according\ Eq.~\eqref{eq6}.~b)~Top\ right.\ 
         Imaginary\ part\ of\ dielectric function\ 
         calculated according to\ Eq.~\eqref{eq5}.\
         c)~Bottom\ left.\ Electron energy loss\ 
         calculated according to\ Eq.~\eqref{eq7}.\
         d)~Bottom right.\ JDOS\ calculated\ 
         according to\ Eq.~\eqref{eq8}.\label{fig6}}

\end{figure*}

We present\ optical properties of pristine\ 
in\ Fig.~\ref{fig4}\ to\ reveal\ differences\ 
as\ compared\ the\ doped\ counterparts\
(Figs.~\ref{fig5}~$\&$~\ref{fig6}).\
It\ looks\ that\ the\ real\ component\ of\ 
the\ dielectric\ function\ has\ non-zero\ 
value\ in\ the\ photon\ energy\ ranges\ of\ 
up\ to\ 3.0~eV.\ {$\varepsilon$}(0)=20,\ and\
${\varepsilon}_{1}$\ approaches\ zero\ at\ 
higher\ photon\ energies\ exceeding\ 4.0~eV.\ 
Conditions\ of\ applicability\ in\ plasmonic\ 
effect\ may\ also\ be\ significant\ at\ 
photon\ energies\ of\ near\ 8.0~eV.\ The\ 
optical\ absorption\ peak\ seems\ to\ 
appear\ near\ photon\ energy\ range\ of\ 
2.0~eV.\ This\ makes\ it\ a\ possible\ 
candidate\ for\ photovoltaic effect\ of\ 
orange and red light.\ Direction\ resolved\ 
optical\ property\ is\ presented\ and\ 
it\ shows\ that\ directional\ interaction\ 
of\ light\ along\ the\ tube\ z-axis\ 
dominates\ the\ contribution\ to\ the\ 
optical\ property.\ JDOS\ shows\ population\ 
of\ holes\ created\ due\ to\ excitation\ 
of\ electrons.\ As\ shown\ in\ 
figure~(bottom right),\ the\ JDOS\ has\ 
significant\ values\ at\ the\ photon\ 
energies\ exceeding\ 1.0~eV,\ with\ peaks\ 
occurring\ in\ the\ photon\ energy\ 
ranges\ of\ 4.0-7.0~eV.\\ 
Optical\ property\ with\ Fe\ doped\ (8,0)\ SiC~SWNT\
is\ presented\ in\ Fig.~\ref{fig5}.\ {$\varepsilon$}(0)=30\
and\ shows\ increases\ compared\ to\ the\ pristine.\ 
Non-zero\ optical\ absorption\ happens\ at\ photon\ 
energies\ from\ 0.3-2.5~eV.\ The\ absorption\ property\ 
at\ photon\ energies\ of\ 0.3-1.5~eV\ occurs\ due\ 
to\ the\ Fe\ doping\ introduced\ to\ the\ system.\ 
At\ these\ absorption\ regime,\ the\ application\ 
as\ detectors\ or\ sensors,\ as\ well\ as\ photovoltaic\
effect\ of\ red\ and\ near\ infrared light\ is\ 
likely\ a\ favourite.\ In\ the\ absorption\ 
regime\ exceeding\ 1.5~eV but up to\ 2.5~eV,\ 
a\ photovoltaic\ effect\ of\ green\ and\ blue\ 
light\ may\ dominate.\ JDOS\ possesses\ non-zero\
values\ beginning\ from\ photon\ energies\ 
exceeding\ 0.2~eV,\ where\ a widening\ peaks\ 
happen\ at\ photon\ energies\ between\ 3.0-7.0~eV.\ 
Energy\ loss\ function\ may\ show\ application\ 
in\ plasmonic\ effect\ near\ photon\ energies\
of\ 8.0~eV.\ 
Optical\ property\ of\ Co\ doped\ (8.0)\ SiC~SWNT\ 
is\ presented\ in\ Fig.~\ref{fig6}.\ The\ properties\ 
investigated\ seem\ to\ be\ closely\ similar\ to\ 
the\ presentation\ of\ the\ case\ of\ Fe\ doped\ 
(8,0)\ SiC~SWNT~(Fig.~\ref{fig5}).\ It\ seems\ that\ 
the\ cobalt\ doping\ could\ result\ to\ significant\ 
photovoltaic\ effects\ of\ green\ and\ blue\ light,\
while\ also\ possibly\ important\ for\ applications\ 
in\ sensors\ and\ detectors.\ Energy\ loss\ function\ 
may\ show\ applications\ in\ plasmonic\ effect\ 
at\ photon\ energies\ near\ 8.0~eV.\  
\subsection{Hydrogen adsorption on pristine and TM doped (8,0)\ SiC~SWNT}
\begin{figure*}[htbp!]
\subfloat a){{\includegraphics[width=.43\textwidth]{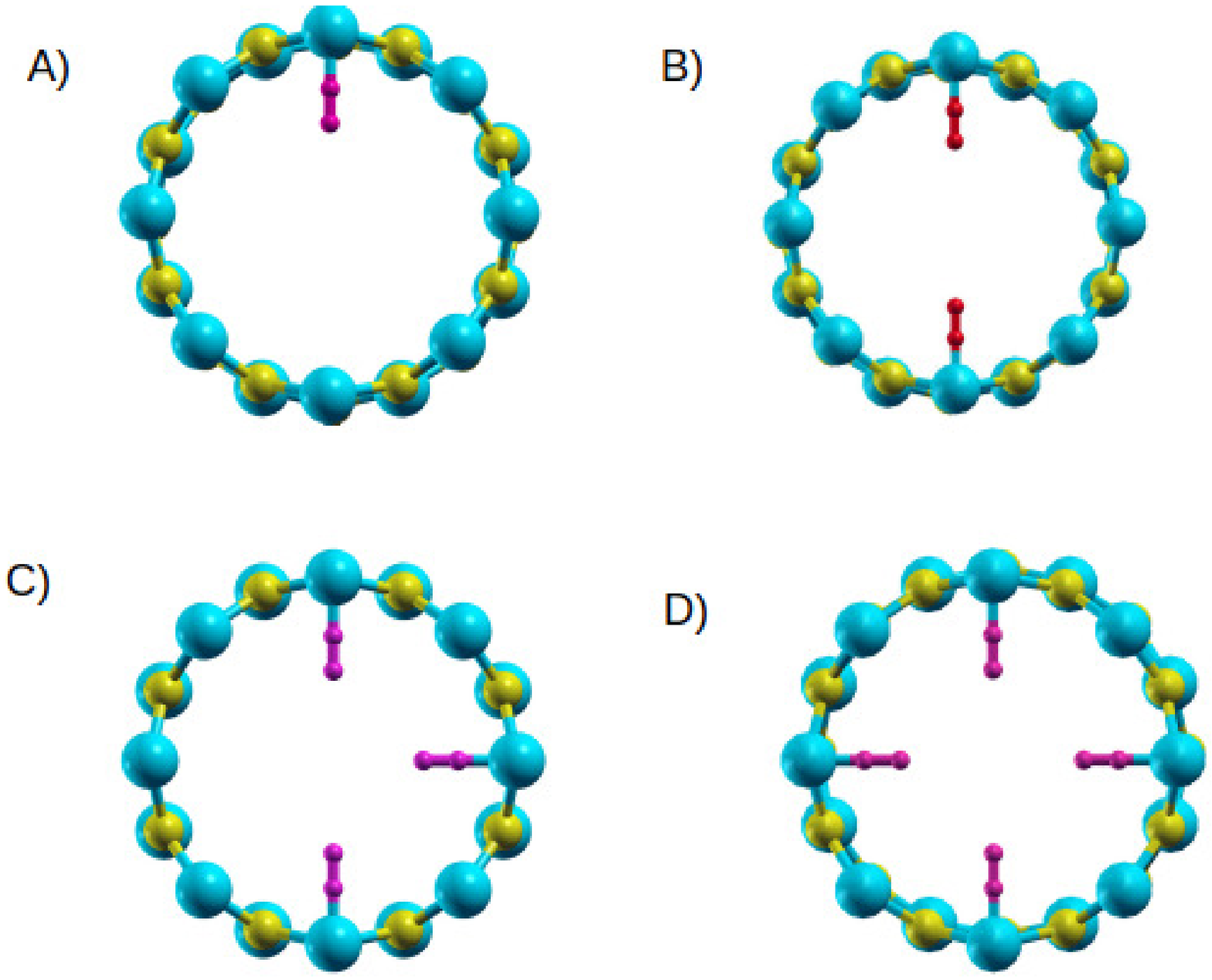}}}\qquad
\subfloat a){{\includegraphics[width=.48\textwidth]{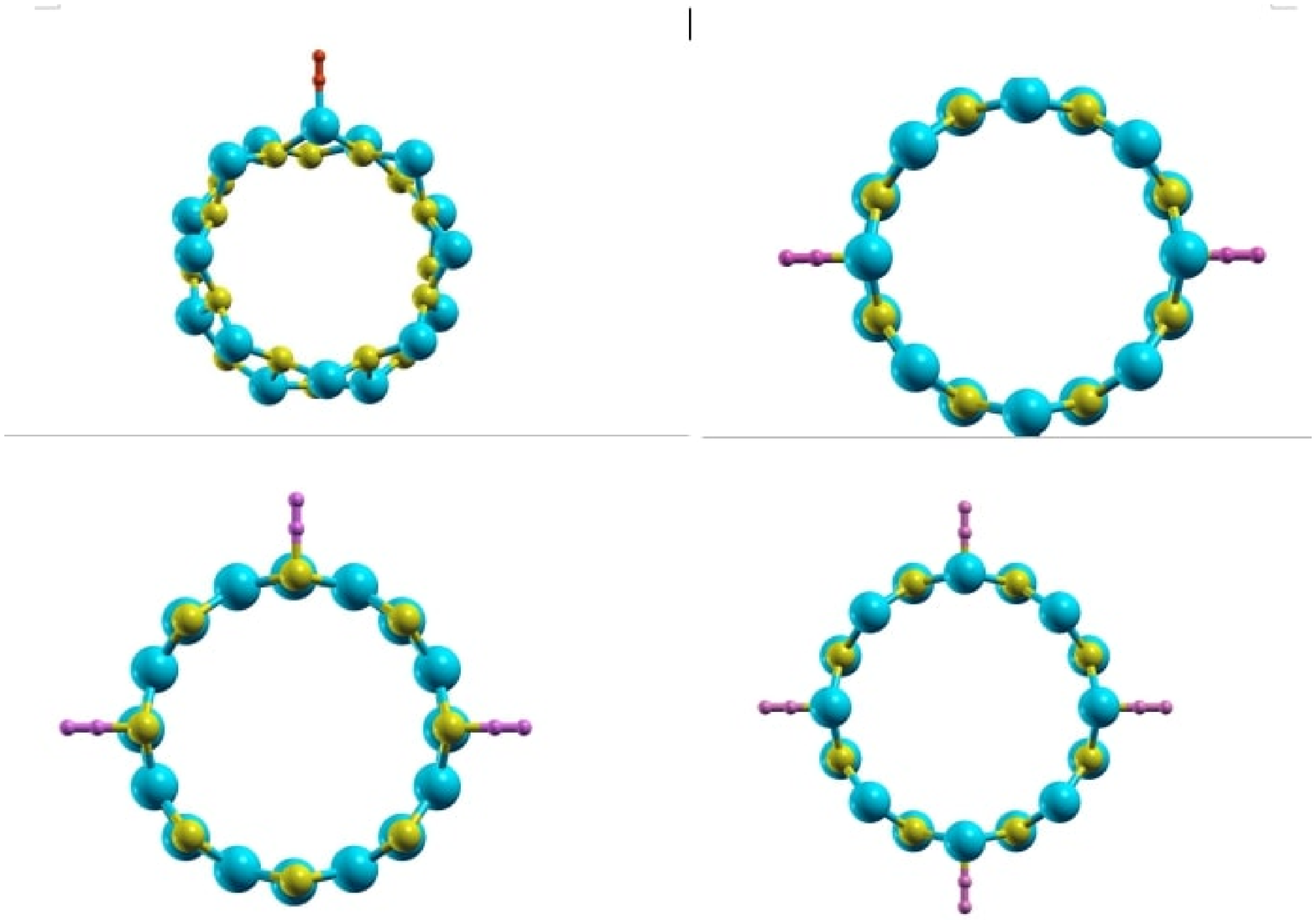}}}\qquad
\subfloat b){{\includegraphics[width=.43\textwidth]{./Figures/Fig7c.eps}}}\qquad
\subfloat c){{\includegraphics[width=.43\textwidth]{./Figures/Fig7d.eps}}}\qquad
\caption{Upper side.\ The\ geometries\ of\ 
         adsorptions\ on\ the\ inside and\ 
         outside\ sites~(see\ table~\ref{tab2}).\ 
         Lower side. Corresponding density\ 
         of states.\label{fig7}}

\end{figure*}

\begin{table}[ht!]
\setlength{\tabcolsep}{10.5mm}
\renewcommand{\arraystretch}{1.5}
\centering
\small
\caption{Formation\ energy~($E\rm_{f}$)~in~eV,\ 
         Total energy per\ atom~($E$)~in~eV/atom,\ 
         Adsorption\ energy~($E\rm_{ads}$)~in~eV,\
         and $n$\ is number of hydrogen molecule
         adsorbed (Eq.~\eqref{eq3})\ 
         on\ pristine\ (8,0)\ SiC~SWNT.\label{tab2}}
\begin{tabular}{lcccc}
\hline\hline 
{Site} & {$n$}  & {$E\rm_{ads}$~(eV)} & {$E\rm_{f}$~(eV/atom)} & {$E$~(eV/atom)}\\
\hline
         & 2  & -1.62 &  1.96  & -9.09  \\
         & 4  & 1.78  &  0.76  & -8.63  \\
$Out$    & 6  & 3.84  &  -1.12 & -8.23   \\
         & 8  & 5.90  &  -4.11 & -7.87  \\\hline
         & 2  &-0.29  &  0.96  & -9.07  \\
$In$     & 4  & 1.67  &  0.85  & -8.62   \\
         & 6  & 3.70  & -1.03  & -8.22  \\
         & 8  & 5.61  & -3.91  & -7.86  \\\hline
\end{tabular}
\end{table} 
From\ table~\ref{tab2},\ it\ looks\ that\ 
adsorption\ energy\ increases\ by\ increasing\ 
the\ concentration\ of\ hydrogen\ molecule.\ 
The\ increases\ happen\ for\ both\ the\ $In$\ 
and\ $Out$\ sites~(Fig.~\ref{fig7}).\ From\ the\
adsorption\ energy\ and\ formation\ energy,\ 
we\ can\ see\ that\ the\ formation\ energy\ 
becomes\ exothermic\ at\ higher\ $n$\ molecules\ 
adsorbed.\ The\ formation\ energy\ shows\
endothermic\ process\ for\ $n$=2 and $n$=4,\ 
while\ an\ exothermic\ process\ for\ 
$n$=6 and $n$=8.\ This\ applies\ to\ both\ 
the\ $In$\ and $Out$\ sites.\ Thus,\ the\ 
formation\ energy\ analysis\ suggests\ 
that\ stability\ increases\ with\ increased\ 
$n$.\ The\ $Out$\ site\ has\ relatively\ 
an\ edge\ over\ the\ $In$\ site\ for\ 
adsorptions.\ The\ adsorption\ energies\ 
seem\ to\ be\ relatively larger\ for\ 
adsorptions\ at\ the\ $Out$\ site\ 
when\ compared\ to\ that\ of\ the\ $In$\
site.\ At\ lower\ adsorbate\ amount\ 
of\ below\ $n$=2,\ it\ seems\ that\ 
the\ $In$\ site\ is\ more\ favourable\ 
adsorption\ site\ compared\ to\ that\ 
of\ the\ $Out$\ site.\ However,\ when\ 
the\ adsorbate\ amount\ exceeds\ $n$=2,\
there\ is\ a\ switch\ in\ the\ 
favorability,\ where\ the\ $Out$\ site\ 
becomes\ more\ favourable\ adsorption\ 
site\ compared\ to\ the\ $In$\ site.\ 
Furthermore,\ when\ the\ amount\ of\ 
adsorbate\ exceeds\ $n$=6,\ the\ 
adsorption\ process\ experiences\ more\
binding\ strength\ to\ the\ tube\ and\
thus\ can\ be\ described\ by\ formation\ 
as\ a\ compound.\ This\ latter\ phenomena\
can\ be\ relevant\ to\ the\ concept\ of\ 
hydrogen\ energy\ storage.\\
 
From\ the\ Fig.~\ref{fig7}~(lower sides),\ it\ seems\ 
that\ the\ adsorption\ may\ not\ change\ the\ electric\ 
property\ of\ the\ pristine\ from\ being\ a\ 
semiconductor.\  
\begin{figure}[htbp!]
\subfloat a){{\includegraphics[width=.45\textwidth]{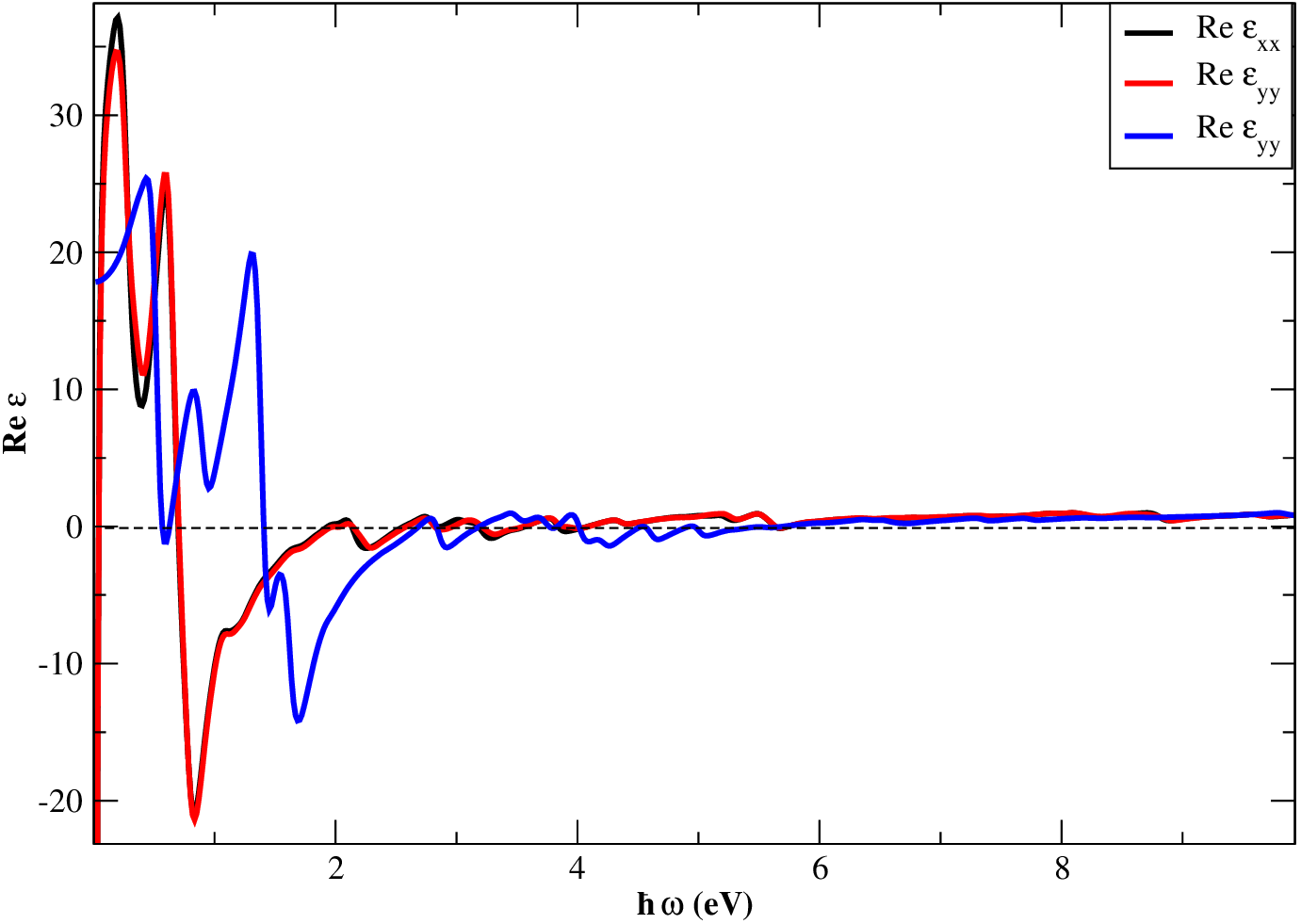}}}\qquad
\subfloat a){{\includegraphics[width=.45\textwidth]{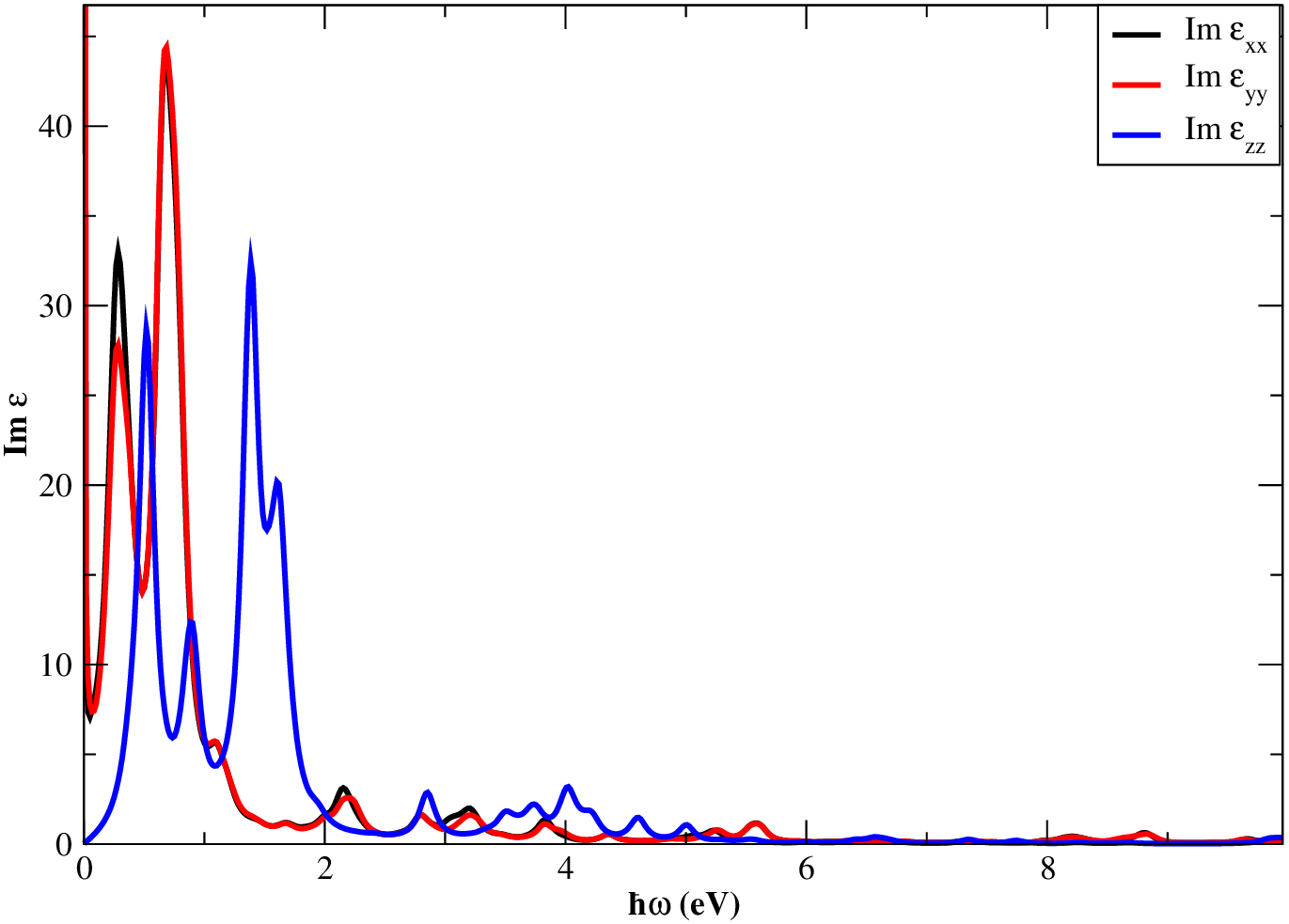}}}\qquad
\subfloat a){{\includegraphics[width=.45\textwidth]{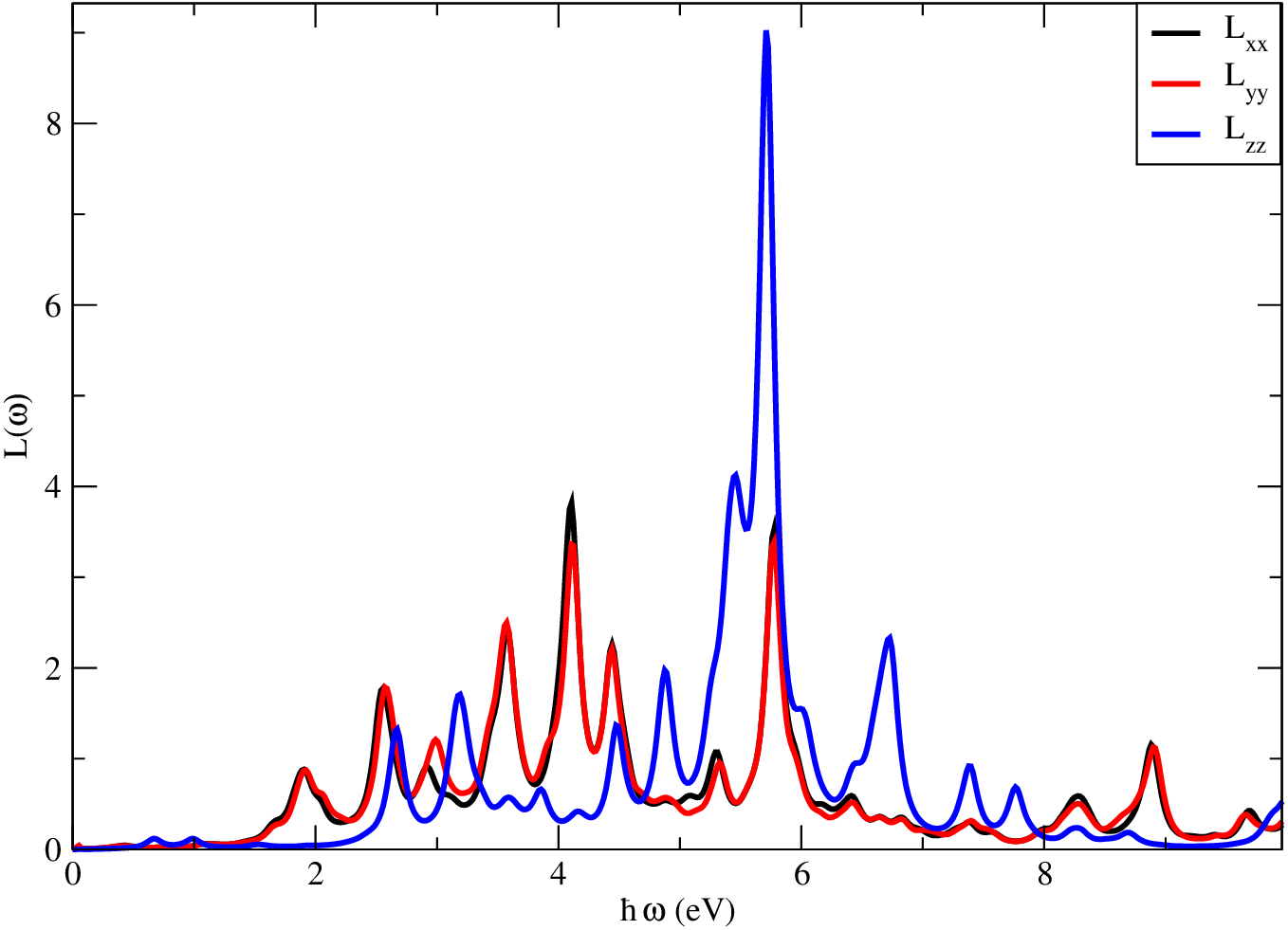}}}\qquad
\subfloat a){{\includegraphics[width=.45\textwidth]{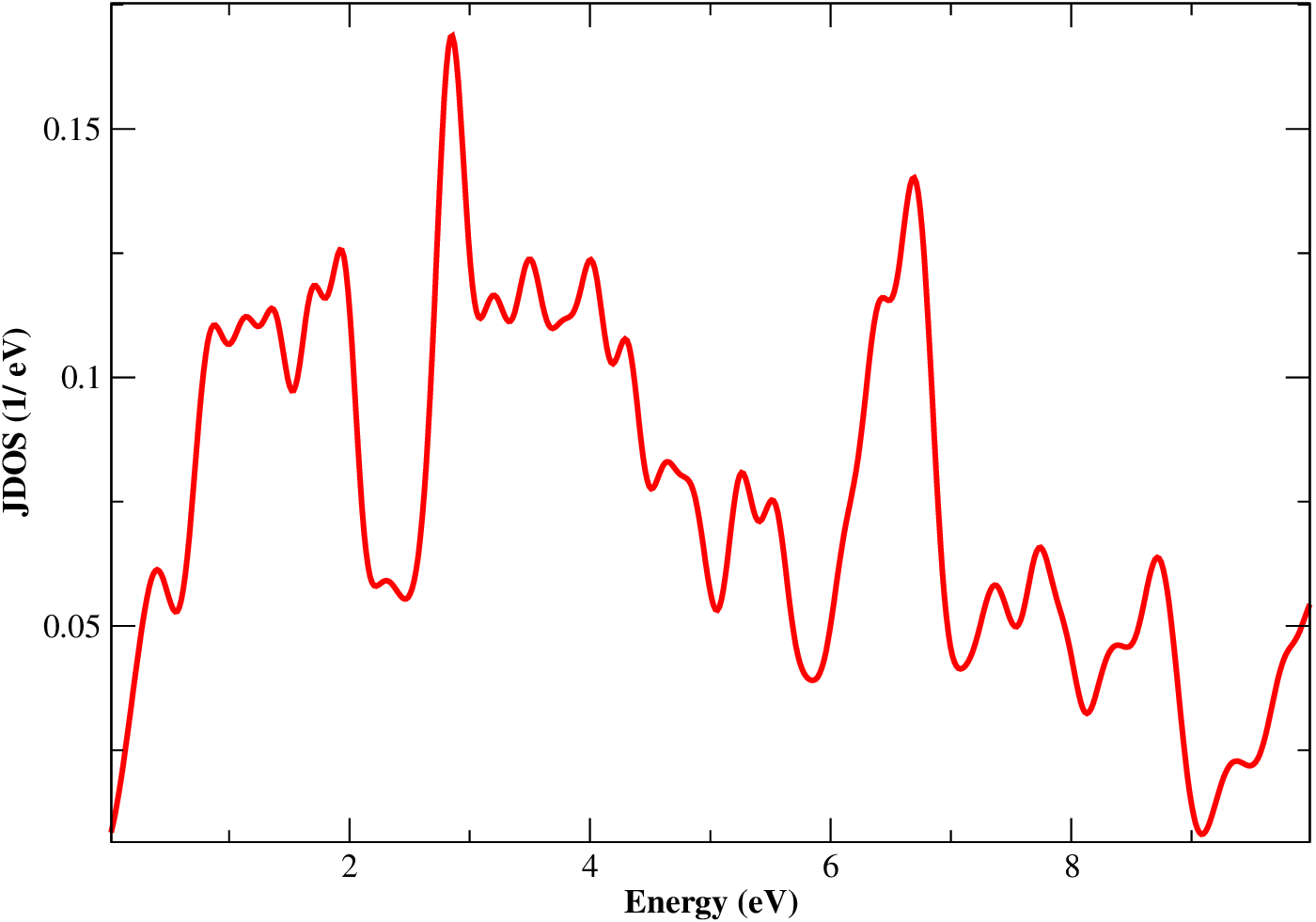}}}\qquad
\caption{Optical\ properties\ of\ pristine\ (8,0)\
         SiC~SWNT\ after hydrogen adsorbed.~a)~Top left.\ 
         Real part\ of\ the\ dielectric function\ calculated\ 
         according\ Eq.~\eqref{eq6}.~b)~Top\ right.\ 
         Imaginary\ part\ of\ dielectric function\ 
         calculated according to\ Eq.~\eqref{eq5}.\
         c)~Bottom\ left.\ Electron energy loss\ 
         calculated according to\ Eq.~\eqref{eq7}.\
         d)~Bottom right.\ JDOS\ calculated\ 
         according to\ Eq.~\eqref{eq8}.\label{fig8}}
\end{figure}
The\ optical\ property\ studied\ following\ 
adsorption\ of\ hydrogen\ is\ given\ in\ 
Fig.~\ref{fig8}.\ By\ adsorption\ of\ 
hydrogen,\ in\ addition\ to\ the\ optical\ 
property\ along\ the\ tube\ z-axis,\ the\ 
optical\ property\ along\ the\ tube's\ y-axis\ 
also\ becomes\ significant.\ As\ a\ result,\ 
applications\ in\ photo-sensors\ and\ 
detectors\ dominate\ in\ photon\ energy\
ranges\ of\ up\ to\ 1.1~eV,\ while\ application\
in\ photovoltaic effect\ of\ near infrared and 
red lights\ dominate in photon\ energies 
range exceeding 1.1~eV\ but\ up\ to\ 2~eV.\ 
Plasmonic\ effects\ can\ potentially\ 
take\ place\ at\ lower\ photon\ energies\ 
such as\ at\ photon\ energies\ of\ around\
5.0~eV,\ compared\ to\ the\ counterpart\
in\ pristine\ or\ TM\ doping.\ 
\section{Conclusion\label{sec:conc}}
The\ typical\ outputs\ of\ this\ study\ 
can\ be\ summarized\ as\ follows.\ 
\begin{itemize}
\item Doping\ with\ Fe,\ Co,\ or\ Mn\ appears\
to\ expand\ the\ range\ of\ optoelectronics\ 
application\ of\ the\ tube.\ Potential\ expanded\ 
applications\ include\ photovoltaic\ 
effects\ in\ a\ broad\ range\ of\ from\ far\ 
infrared\ light\ to\ visible\ light.\ 
\item Doping\ of\ Fe\ and\ Mn\ have\ the\ 
potential\ to\ lead\ to\ half-metallic\ 
property\ with\ a\ possible\ application\ 
in\ spintronics.\ 
\item Hydrogen\ storage\ seems\ to\ have\ 
a\ good\ promising\ opportunity\ on\ (8,0)\ 
SiC~SWNT.\ This\ pinpoints\ its\ potential\ 
relevance\ to\ the\ hydrogen\ energy\ storage\
application.\ 
\item Adsorption\ of\ hydrogen\ seems\ to\ 
impact\ on\ optical\ property\ of\ the\ tube\ 
by\ widening\ the\ scope\ of\ application\ 
in\ optoelectronics\ as\ similar\ with\ the\
impacts\ that\ the\ dopings\ offer.\ 
This\ effect\ can\ be\ particularly\ noticeable\
when\ the\ amount\ of\ adsorbed\ hydrogen\
($n$H$\rm_{2}$)\ is\ large,\ i.e.,\ where\ 
$n${$\geq$}6.\ The\ adsorbed\ hydrogen\ prefers\
to\ interact\ with\ silicon\ atoms\ of\ the\ 
tube\ than\ carbon\ atoms.\ Furthermore,\ 
the\ adsorbates\ bind\ to\ the\ tube\ 
preferably\ on\ the\ outside\ of\ the\ 
tube\ than\ on\ the\ inside\ of\ the\ 
tube.\ 
\item Direction\ resolved\ analysis\ of\ 
the\ interaction\ of\ light\ with\ the\ 
tube\ indicates\ that\ the\ tube\ as\ a\ 
pristine\ or\ in\ a\ doped\ status\ dominantly\
interacts\ along\ its\ periodic\ z-axis.\ 
But\ in\ the\ case\ of\ hydrogen\ being\ 
already\ adsorbed,\ significant\ interaction\
along\ the\ tube's non-periodic\ y-axis\
happens\ in\ addition\ to\ the\ periodic\
z-axis.\ 
\item The\ tube\ in\ its\ pristine or doped\ 
status\ appears\ to\ have\ significance\ 
for\ plasmonic effect and photonics applications\
at\ photon\ energies\ in\ the\ far\ ultraviolet\
light\ (i.e., ca\ ${\hbar}{\omega}$=8~eV).\ 
\end{itemize}

\section*{Disclosure\ statement}
The authors declare\ that there is no conflict\ 
of interest.

\section*{Acknowledgments}
We are grateful to\ the Ministry of\ Education\ 
of Ethiopia\ for~financial~support.~The authors\ 
also acknowledge~the\ International Science\ 
Program,~Uppsala~University,~Sweden,\ for\ 
providing computer~facilities~for~research~at~the\ 
Department~of~Physics.~The~office~of~VPRTT~of Addis\ 
Ababa\ university is also warmly~appreciated~for\ 
supporting~this\ research under a\ 
grant~number~AR/032/2021.\ 
\section*{ORCID\ iDs}
K.N.\ Nigussa.\\
\url{https://orcid.org/0000-0002-0065-4325}.
\section*{References}
\bibliographystyle{elsarticle-num}
\bibliography{arXrefs.bib}
\end{document}